# Measuring Transition Risk in Investment Funds

Ricardo Crisóstomo[a]

## Abstract

We develop a comprehensive framework to measure the impact of the climate transition on investment portfolios. Our analysis is enriched by including geographical, sectoral, company and ISIN-level data to assess transition risk. We find that investment funds suffer a moderate 5.7% loss upon materialization of a high transition risk scenario. However, the risk distribution is significantly left-skewed, with the worst 1% funds experiencing an average loss of 21.3%. In terms of asset classes, equities are the worst performers (-12.7%), followed by corporate bonds (-5.6%) and government bonds (-4.8%). We discriminate among financial instruments by considering the carbon footprint of specific counterparties and the credit rating, duration, convexity and volatility of individual exposures. We find that sustainable funds are less exposed to transition risk and perform better than the overall fund sector in the low-carbon transition, validating their choice as green investments.

**Keywords:** Climate change. Low-carbon transition. Asset allocation. Investment funds. NGFS scenarios.

[a] Comisión Nacional del Mercado de Valores (CNMV), Edison 4, 28006 Madrid, Spain. The opinions in this paper are solely those of the author and do not necessarily reflect those of the CNMV.
Email: rcayala@cnmv.es

# 1. Introduction

The transition towards a low-carbon economy might trigger financial stability concerns stemming from the materialization of transition scenarios than are not anticipated by economic agents. Changes in investors' preferences, technological disruptions and the abrupt implementation of climate policies introduce a future uncertainty that can cause a sudden revaluation of financial assets. The green transition is expected to increase the costs of carbon-intensive firms while decreasing the demand for their products. These changes could generate stranded assets, deteriorated credit quality, reduced firm valuations and higher financing costs, leading to losses in the financial instruments issued by transition-vulnerable firms.

Analyzing the links between investment funds and high-carbon firms can provide early warning indicators of the systemic risk originating from the green transition. Higher climate awareness and the disclosure of new environmental information could make investors to reduce or reject high-carbon investments, generating contagion across overlapping exposures and a risk of runs on brown assets (Jondeau et al., 2021a; Jondeau et al., 2021b)[1]. Beyond the direct effect on brown assets, financial institutions and investors with ownership and debt links to high-carbon firms could be also affected by the transition through increased credit and market risk in their portfolios.

This paper develops a comprehensive framework to measure transition risk in investment portfolios. The vulnerability of financial counterparties to the climate transition is derived from an assessment of their carbon intensity and economic sector. Furthermore, market and credit risk metrics are specifically retrieved for each portfolio exposure to account for the varying degree of risk of different financial instruments and asset classes. This methodology allows us to quantify the loss of value that each individual exposure, and hence the corresponding portfolio, could suffer upon materialization of a transition risk scenario.

To our knowledge, there are two papers that employ a related approach to deal with climate risk in investment funds. Amzallag (2021) studies the climate risk of investment funds by means of their carbon footprint and shows that investment fund exposures to the climate transition are heterogeneous, with funds investing in highly pollutant firms exhibiting higher interconnectedness than funds investing in sustainable activities. Gourdel and Sydow (2021) analyze the impact of physical and transition risks on European funds considering redemption shocks, losses from repricing risk, fire sales and second-round effects. Their result shows a better performance of green funds in the climate transition but generalized losses in case of a physical shock.

Compared to the literature, our contribution is threefold:

- We introduce an ISIN-level methodology that can be used to quantify the climate risk of any investment portfolio. Our approach considers both climate and financial risk metrics to measure transition risk.

- We perform the first climate-related description of the Spanish investment fund universe. We derive from this analysis the main characteristics of Spanish fund portfolios, including (i) a comparison with their European peers and (ii) an analysis of how sustainable funds perform in the climate transition.

---

[1] Contagion can occur when a fund selling its portfolio cause losses to other market participants with overlapping exposures (Poledna et al., 2021)

- The granularity of our analyses is enriched by considering sectoral, geographical, company and ISIN-level data. Compared to sectoral models, we find that the product mix, energy reliance and technological portfolio of specific companies can significantly alter their climate risk profile.

Our analyses show that investment funds suffer moderate losses upon a materialization of a high transition risk scenario. Overall, the average mark-to-market (MtM) loss in the investment funds sector is 5.69%. However, the distribution of transition risk losses is significantly left-skewed, with the 1% worst performing funds enduring an average MtM loss of 21.34%. While many funds respond resiliently to the low-carbon transition, funds' investing in equities of highly pollutant companies suffer the highest losses. These figures should be interpreted as a low severity estimate of the potential fund losses, as only direct portfolio effects are considered. Amplifying factors like the fund-flow relation, market impact, indirect contagion or other system-wide drivers could trigger feedback loops and non-linear effects that increase the final loss[2].

We also find that sustainable funds outperform the overall fund sector in the climate transition. In terms of tail risk, although sustainable funds held a higher share of equity investments, the worst 1% and 5% vehicles suffer a limited 14.65% and 11.00% loss (vs 21.34% and 15.47% in the fund sector). Similarly, the MtM loss observed across all sustainable funds is 5.70%, which is lower than the 5.92% attributed to the comparable portfolio in terms of asset classes. These figures indicate that sustainable funds are less exposed to transition risk and invest in financial assets that outperform their sectoral peers in the low-carbon transition.

Finally, we show that Spanish funds exhibit lower transition risk than their European counterparts. Using the framework developed by Alessi and Battiston (2021), the Transition-risk Exposure Coefficient (TEC) of Spanish funds is 4.37% versus 6.11% in EU funds. Considering the portfolio share that is included in the calculation, the adjusted TEC of Spanish funds increase to 12.91% (vs. 29.2% in EU funds). Furthermore, sustainable fund portfolios exhibit significantly lower TEC and adjusted TEC than both Spanish and EU funds, reinforcing their choice as green investments.

The remainder of the paper is structured as follows: Section 2 introduces the methodology employed to assess transition risk in investment portfolios. Section 3 presents the data employed and the climate scenario considered. Section 4 shows the results from our transition risk analyses. Finally, Section 5 concludes.

---

[22] See for instance, Clerc et al. (2016), Peralta and Crisóstomo (2016), Cont and Schaanning (2017), Battiston et al. (2017), Ojea-Ferreiro (2020), Roncoroni et al. (2021), or Alessi et al. (2022).

## 2. Measuring transition risk in investment portfolios

We develop a comprehensive framework to assess the vulnerability of investment portfolios to the low-carbon transition. Analytically, we consider both climate indicators and traditional measures of credit and market risk. Figure 1 provides a schematic description of the steps and risk factors employed to quantify transition risk.

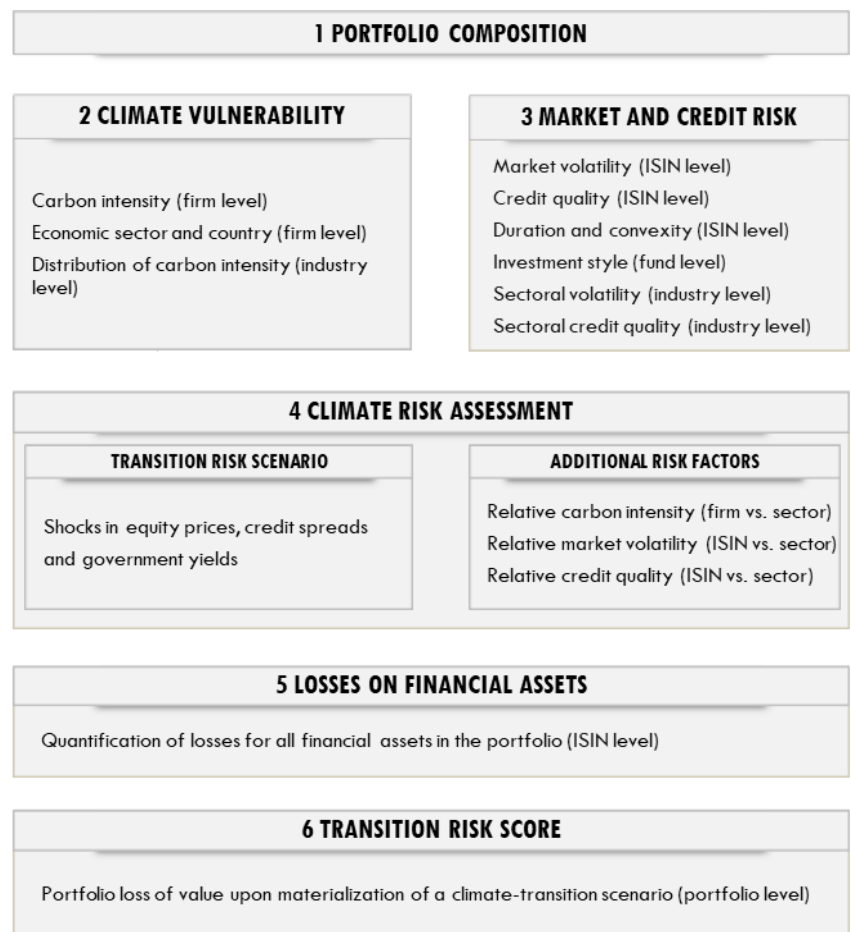

**Figure 1:** Diagram illustrating the steps and risk factors used to quantify transition risk.

Our climate risk assessment starts with the ISIN-level composition of financial portfolios. The vulnerability of each counterparty to the climate transition is derived from their carbon intensity and economic sector of operation. Companies operating with high carbon footprints are expected to be more affected by the green transition, as higher carbon prices, changes in investor's preferences, technological disruptions and climate-related policies could reduce the demand for carbon intensive products while increasing production costs. Consequently, the green transition is expected to generate stranded assets, reduced firm valuations and increased credit risk, leading to losses in the financial instruments issued by high-carbon firms.

In addition, the characteristics of economic sectors constitute a relevant risk driver to assess the cost and opportunities arising from the climate transition. Economic sectors with higher levels of GHG emissions, such as utilities, transportation, mining and petroleum are most at risk of suffering climate-related losses, as policies aimed at curbing emissions and facilitating the green transition create significant risks to carbon-intensive industries (see UNEP FI, 2019; BCBS, 2021; Dunz et al., 2021 and Semieniuk et al., 2021)

However, even within a given industry, climate risk usually affects companies heterogeneously. Company-specific drivers like their product mix, reliance in different energy sources or technological portfolio can significantly alter the climate risk profile of individual firms. Therefore, in addition to sectoral aggregates, we employ company-level data to perform intra-sectors comparisons and discriminate among the best and worst-in-class in each economic industry (see the CO-Firm and Kepler Cheuvreux, 2018 and NGFS, 2020).

Furthermore, even for a given company, the MtM performance of financial instruments will be typically different depending on the asset class considered and the risk characteristics of specific exposures. For credit instruments, the risk of loss can differ depending on the credit quality, collateralization, duration, and convexity of each exposure. Similarly, the risk of loss in equity instruments also varies depending on market risk factors like the volatility of each underlying.

Building on this, we develop a portfolio methodology that comprehensively assesses the climate risk of five interrelated asset classes: i) equity investments; ii) corporate bonds, iii) sovereign debt; iv) investment in other fund vehicles; and v) cash and cash-equivalents.

## 2.1 Equity investments

The climate vulnerability of equity exposures is derived from the economic sector and carbon intensity of each counterparty. For equity instruments, we first assess the carbon intensity of specific firms relative to its sector as:

$$CI_{i,j}^m = 2q_j(CI_i)$$

where $CI_{i,j}^m$ represents the carbon intensity multiplier of exposure i, j and $q_j(CI_i)$ is the quantile that the issuer of i occupies in the carbon intensity distribution of sector j. For each counterparty, $CI_{i,j}^m$ ranges from 0 to 2, effectively discriminating among firms that deviates from the median carbon intensity of its economic sector.

We next obtain the market risk multiplier of equity investments by comparing the volatility of each exposure $\sigma_i$ with the average volatility of its economic sector $\bar{\sigma}_j$:

$$\sigma_{i,j}^m = \frac{\sigma_i}{\bar{\sigma}_j}$$

After assessing the climate vulnerability and market risk, the loss of value in equity position i, j is obtained as

$$\Delta \text{EQ}_{i,j}(k) = CI_{i,j}^m \, \sigma_{i,j}^m \, \Delta_j^{EQ}(k)$$

where $\Delta_j^{EQ}$ represents the average equity loss in sector j upon materialization of transition scenario k.

## 2.2 Corporate bonds

Similar to equity investments, the climate vulnerability of corporate debt is derived from the economic sector and carbon intensity of each counterparty. The carbon intensity multiplier of corporate bond i pertaining to sector j is

$$CI_{i,j}^m = 2q_j(CI_i)$$

where $q_j(CI_i)$ is the quantile that the obligor of debt i occupies in the carbon intensity distribution of sector j. Next, we obtain the credit risk multiplier through the Credit Quality Step (CQS) of each individual exposure. The CQS provides a standardized mapping of credit ratings that considers risk drivers from both the obligor (e.g.: probability of default) and the exposure (e.g.: subordination, expected recovery, existence of guarantees/collateral, etc.)[3]. For each corporate debt, we calculate the CQS multiplier as:

$$CQS_{i,j}^m = \frac{CQS_i}{\overline{CQS}_j}$$

where $CQS_i$ is the credit quality step of debt i, and $\overline{CQS}_j$ is the average CQS of economic sector j. Table 1 provides the mapping between the CQS and the credit ratings of the main agencies.

Table 1: Mapping of credit ratings to credit quality steps

| CQS | Fitch's ratings | Moody's ratings | S&P's ratings |
|---|---|---|---|
| 1 | AAA to AA- | Aaa to Aa3 | AAA to AA- |
| 2 | A+ to A- | A1 to A3 | A+ to A- |
| 3 | BBB+ to BBB- | Baa1 to Baa3 | BBB+ to BBB- |
| 4 | BB+ to BB- | Ba1 to Ba3 | BB+ to BB- |
| 5 | B+ to B- | B1 to B3 | B+ to B- |
| 6 | CCC+ and below | Caa1 and below | CCC+ and below |

After assessing credit risk and climate vulnerability, the credit spread change of corporate debt i, j is obtained as

$$\Delta CS_{i,j}(k) = CI_{i,j}^m \, CQS_{i,j}^m \, \Delta_j^{CS}(k)$$

where $\Delta_j^{CS}(k)$ is the average credit spread variation of sector j due to transition scenario k. Finally, the loss of value for each corporate bond is obtained as:

$$\Delta CB_{i,j}(k) = \Delta CS_{i,j}(k) SD_i + 0{,}5\Delta CS_{i,j}(k)^2 C_i$$

where $SD_i$ and $C_i$ represent the credit spread duration and the convexity of bond i, respectively.

---

[3] The CQS is a standardized indicator of credit risk derived from the Capital Requirements Regulation (see Regulation EU No 575/2013 and Implementing Regulation EU 2016/1799). We reckon that forward-looking metrics (e.g.: credit spread and implied volatilities) are preferable to historical-based measures (see, for instance, Crisóstomo and Couso, 2018 and Wei et al., 2022). However, since forward-looking metrics are only available for a small subset of financial instruments, using credit spreads would significantly reduce data availability in our dataset.

## 2.3 Sovereign debt

The climate vulnerability of sovereign debt is estimated separately for each country. For countries that are explicitly included in the transition pathway k, the change in value in the sovereign debt i of country h is obtained as

$$\Delta GB_{i,h,T}(k) = \Delta y_{h,T}(k) D_i + 0.5 \Delta y_{h,T}^2(k) C_i$$

where $\Delta y_{h,T}(k)$ represents the yield change in the debt of country h and maturity T, whereas $D_i$ and $C_i$ are the modified duration and convexity of sovereign debt i, respectively.

For countries that are not included in the transition pathway k, we estimate the vulnerability of country h from its carbon intensity multiplier

$$CI_{i,h}^m = 2q^{GB}(CI_i)$$

where $q^{GB}(CI_i)$ represents the quantile of the sovereign debt distribution in which country h is located. Next, the credit risk multiplier is obtained through the CQS of government debt i compared with the average CQS of sovereign bond exposures

$$CQS_{i,h}^m = \frac{CQS_i}{CQS^{GB}}$$

And the yield change in exposure i of country h is hence given by

$$\Delta y_{i,h,T}(k) = CI_{i,h}^m \, CQS_{i,h}^m \, \Delta y_T(k)$$

where $\Delta y_T(k)$ represent the average yield change in the sovereign debt of maturity T. Finally, the change in value in government exposure i, h, T is then

$$\Delta GB_{i,h,T}(k) = \Delta y_{i,h}(k) D_i + 0.5 \Delta y_{h,T}^2(k) C_i$$

## 2.4 Other fund vehicles

The transition risk of other fund vehicles is derived from the carbon intensity and composition of each fund portfolio. Similar to other asset classes, we first compare the carbon intensity of fund i with the carbon intensity distribution of the other funds sector as

$$CI_i^m = q^{IF}(CI_i)$$

where $q^{IF}(CI_i)$ represent the quartile of the distribution in which fund i is located. In absence of ISIN-level information, the portfolio composition of other fund vehicles is derived from the investment style of each fund, as shown in Table 2.

Next, the change in value in fund i of investment style *s* is obtained as

$$\Delta IF_{i,s}(k) = CI_i^m \left[ w_s^{EQ} \overline{\Delta EQ}(k) + w_s^{CB} \overline{\Delta CB}(k) + w_s^{GB} \overline{\Delta GB}(k) \right]$$

where $w_s^{EQ}$, $w_s^{CB}$ and $w_s^{GB}$ are the share of the portfolio that each fund held in equities, corporate bonds and sovereign debt respectively, whereas $\overline{\Delta EQ}(k)$, $\overline{\Delta CB}(k)$ and $\overline{\Delta GB}(k)$ represent the average change in value of equities, corporate bond and sovereign debt in transition scenario k.

Table 2: Other funds vehicles portfolio by investment style

| Investment style | Equity | Corporate debt | Sovereign debt | Cash and cash equivalents |
|---|---|---|---|---|
| Equities | 85% | 5% | 5% | 5% |
| Mixed equities | 65% | 15% | 15% | 5% |
| Mixed bonds | 25% | 35% | 35% | 5% |
| Bonds | 0% | 75% | 20% | 5% |
| Government debt | 0% | 20% | 75% | 5% |
| Others | 25% | 35% | 35% | 5% |

## 2.5 Cash and cash-equivalents

Cash and cash-equivalents are considered safe assets and hence are not expected to lose MtM value in as adverse climate scenario. As a result, investment funds with higher cash investments will, ceteris paribus, exhibit lower losses upon materialization of an adverse transition risk scenario.

## 3. Data and calibration

Individual fund data is obtained from the position-by-position (i.e.: ISIN-level) portfolio reported by investment funds to the CNMV. We retrieve the portfolio composition for all funds as of June 2021. Our database is composed by 1,629 investment funds with 88,631 individual positions. Fund exposures are categorized into five asset classes that collectively represent over 99% of the assets under management (AuM) of Spanish investment funds[4]: i) Equities; ii) Corporate bonds; iii) Sovereign debt; iv) Investment in other fund vehicles; and v) Cash and cash-equivalents. The total AuM in our database is EUR 307.4 billion. Table 3 show the composition of investment funds portfolios by asset class.

Table 3: Investment funds portfolio by asset class

| Asset class | Investment share (AuM %) | N. of positions | Unique ISINs |
|---|---|---|---|
| Equity | 15.46% | 31,834 | 4,196 |
| Corporate bonds | 19.68% | 28,274 | 5,598 |
| Sovereign debt | 20.97% | 8,532 | 1,462 |
| Other fund vehicles | 34.42% | 12,877 | 3,802 |
| Cash and cash-equivalents | 8.81% | 6,191 | - |
| Not classified | 0.65% | 923 | - |

### 3.1 Data for individual positions

Our transition risk methodology assesses the vulnerability of all fund exposures to the low-carbon transition. The climate sensitivity is assessed at the counterparty level whereas market and credit risk indicators are retrieved for each ISIN-level position. Table 4 summarize the climate and financial inputs retrieved for each asset class, the AuM coverage, and the data sources.

To obtain the climate and financial inputs, we start with information available from the ISIN of each position. Market and credit risk metrics (i.e.: CQS, duration, convexity, volatility and investment style) are obtained for each ISIN code, whereas climate-related metrics (i.e.: carbon intensity, economic sector and country) are retrieved from the obligor of each exposure. When a climate risk metric is not available at the obligor level, we retrieve the information from the parent company or ultimate parent company. This procedure yields an AuM coverage of 97.1% on average, ranging from 91.2% to 100% depending on the data field. For the small number of exposures where carbon intensity, CQS or volatility data could not be retrieved, we employ a sector-neutral backfill.[5]

---

[4] The remainder represents the individual positions that could not classified due to lack of data in the reported portfolio. Derivative positions were not included in our assessment but will be covered in future assessments. Fund compartments with a separate portfolio and investment strategy (i.e.: sub-funds) are considered individual funds in our risk assessment.

[5] Consequently, when the carbon intensity, CQS or volatility data is not available, our backfill method result in a neutral carbon intensity, CQS or volatility multiplier of 1.

**Table 4: Climate and financial risk metrics by asset class**

| Asset class | Climate risk metrics | | | Financial risk metrics | | | | |
|---|---|---|---|---|---|---|---|---|
| | Carbon intensity | Economic sector | Country | Credit quality step | Duration | Convexity | Volatility | Investment style |
| Corporate bonds | ✓ | ✓ | - | ✓ | ✓ | ✓ | - | - |
| Sovereign debt | ✓ | - | ✓ | ✓ | ✓ | ✓ | - | - |
| Equities | ✓ | ✓ | - | - | - | - | ✓ | - |
| Other funds | ✓ | - | - | - | - | - | - | ✓ |
| AuM coverage (%) | 91.8 | 100 | 100 | 93.5 | 100 | 100 | 91.2 | 100 |
| Data sources | Refinitiv, Bloomberg, MSCI, own calc. | Refinitiv, own calculations | Refinitiv, own calculations | Refinitiv, own calculations | Refinitiv, own calculations | Refinitiv, own calculations | Refinitiv, own calculations | MSCI, own calculations |

### 3.2 Climate risk scenario

To ensure a consistent risk assessment across economic sectors, geographical breakdowns, asset classes and individual exposures, we employ a top-down modelling approach. Our analyses start with a global climate scenario that generates macroeconomic and environmental projections with sectoral and geographical breakdown. Next, to operationalize this scenario, we employ a downscaling model that considers both climate and financial risk metrics to obtain a comprehensive risk assessment for different asset classes, counterparties and individual financial instruments.

In the financial sector, the scenarios developed by the Network for Greening the Financial System (NGFS) provide a common reference framework for analyzing climate risks to the economy and the financial system. Regarding transition risk, one of the NGFS scenarios assumes a delayed transition in which climate policies are not introduced until 2030. Consequently, an abrupt (disorderly) implementation is required in 2030 to limit global warming, concentrating the effects of the climate transition in a short period. The delayed transition scenario prompts a rapid increase in carbon prices that generates geographical and sectoral shocks which affect the overall economy (see NGFS, 2021).

The macroeconomic and environmental projections consistent with the NGFS delayed transition are obtained from the NiGEM and REMING-MagPIE models[6]. Using these projections, the ESRB and ECB provide climate-related shocks for different asset classes that are representative of the delayed transition. Table 5 shows the shocks in equities and credit spreads by economic sector, whereas table 6 exhibits the shocks in sovereign debt yields by country. These figures concentrate the impact in asset prices that are expected over a 3-year period (2030-33) and have been employed in the 2022 stress test for EU pension funds (see ESRB, 2022 and EIOPA, 2022). The sectoral disaggregation included in table 5 is also used in the 2022 ECB climate stress for banks (ECB, 2022).

---

[6] NIGEM is a transparent, peer-reviewed, global macro econometric model that is developed and maintained by National Institute of Economic and Social Research (see UNEP FI, 2022). The REMIND-MAgPIE framework blends the energy-economy model REMIND and the agricultural production model MAgPIE and is developed and maintained by the Potsdam Institute for climate impact research (see Hilaire and Bertram, 2020).

**Table 5: Shocks in equity prices and corporate bond spreads**

| NACE codes | Equity prices (% change) | Corporate bond spreads (basis points change) | Sector description |
|---|---|---|---|
| A01 | -11.50% | 143 | Crop, animal production, hunting and related services |
| A02-A03 | -11.80% | 146 | Forestry, logging, fishing and aquaculture |
| B05-B09 | -37.80% | 467 | Mining and quarrying |
| C10-C12 | -12.30% | 152 | Manufacture of food products, beverages and tobacco |
| C13-C18 | -10.90% | 134 | Manufacture of textiles, wearing apparel, leather, paper and related products |
| C19 | -32.20% | 397 | Manufacture of coke and refined petroleum products |
| C20 | -12.70% | 157 | Manufacture of chemicals and chemical products |
| C21-C22 | -11.10% | 137 | Manufacture of pharmaceutical products and preparations, rubber and plastic products |
| C23 | -20.40% | 252 | Manufacture of other non-metallic mineral products |
| C24-C25 | -15.30% | 189 | Manufacture of basic and fabricated metal products, except machinery and equipment |
| C26-C28 | -11.10% | 138 | Manufacture of computer, electronic, optical, electrical equipment and machinery |
| C29-C30 | -11.20% | 139 | Manufacture of motor vehicles, trailers and semi-trailers and other transport equipment |
| C31-C33 | -9.80% | 121 | Manufacture of furniture and others. Repair and installation of machinery and equipment |
| D35 | -23.00% | 284 | Electricity, gas, steam and air conditioning supply |
| E36-E39 | -13.10% | 162 | Water supply; sewerage; waste management and remediation activities |
| F41-F43 | -11.50% | 143 | Construction |
| G45-G47 | -13.40% | 165 | Wholesale and retail trade; repair of motor vehicles and motorcycles |
| H49 | -22.60% | 279 | Land transport and transport via pipelines |
| H50 | -12.70% | 157 | Air transport |
| H51 | -14.20% | 176 | Water transport |
| H52-H53 | -10.80% | 133 | Warehousing and support activities for transportation; Postal and courier activities |
| L68 | -12.00% | 148 | Real estate activities |
| Other | -14.30% | 177 | Other activities |

Source: ESRB(2022) and EIOPA(2022).

**Table 6: Shocks in sovereign debt yields**

| Country | Sovereign bond yields (basis points change) | | | | | | | | | |
|---|---|---|---|---|---|---|---|---|---|---|
| | 1y | 2y | 3y | 4y | 5y | 6y | 7y | 8y | 9y | 10+y |
| Austria | -7.8 | 6.2 | 20.2 | 34.2 | 48.2 | 62.2 | 76.2 | 90.1 | 104.1 | 118.1 |
| Belgium | 85.7 | 89.7 | 93.6 | 97.6 | 101.6 | 105.6 | 109.6 | 113.6 | 117.6 | 121.6 |
| Bulgaria | 111.9 | 112.2 | 112.5 | 112.8 | 113.1 | 113.4 | 113.8 | 114.1 | 114.4 | 114.7 |
| China | -63.2 | -45.6 | -27.9 | -10.3 | 7.4 | 25.1 | 42.7 | 60.4 | 78.0 | 95.7 |
| Croatia | 316.2 | 289.9 | 263.6 | 237.3 | 211.1 | 184.8 | 158.5 | 132.2 | 106.0 | 79.7 |
| Cyprus | 78.5 | 80.1 | 81.8 | 83.4 | 85.0 | 86.6 | 88.3 | 89.9 | 91.5 | 93.1 |
| Czech Republic | 20.9 | 12.5 | 4.2 | -4.2 | -12.6 | -21.0 | -29.4 | -37.8 | -46.1 | -54.5 |
| Denmark | 93.4 | 95.8 | 98.2 | 100.6 | 103.0 | 105.4 | 107.8 | 110.2 | 112.6 | 115.0 |
| Estonia | 78.4 | 82.5 | 86.5 | 90.6 | 94.6 | 98.7 | 102.7 | 106.8 | 110.9 | 114.9 |
| Finland | 89.0 | 92.4 | 95.8 | 99.2 | 102.7 | 106.1 | 109.5 | 112.9 | 116.3 | 119.8 |
| France | 91.1 | 94.4 | 97.7 | 101.0 | 104.3 | 107.6 | 110.9 | 114.2 | 117.5 | 120.8 |
| Germany | 80.5 | 84.9 | 89.3 | 93.7 | 98.1 | 102.5 | 106.9 | 111.3 | 115.7 | 120.0 |
| Greece | 61.1 | 64.6 | 68.2 | 71.7 | 75.3 | 78.8 | 82.4 | 85.9 | 89.5 | 93.0 |
| Hungary | 74.8 | 54.7 | 34.5 | 14.4 | -5.7 | -25.9 | -46.0 | -66.2 | -86.3 | -106.4 |
| Iceland | -6.3 | -2.4 | 1.5 | 5.5 | 9.4 | 13.3 | 17.3 | 21.2 | 25.1 | 29.0 |
| Ireland | 69.5 | 72.0 | 74.4 | 76.8 | 79.2 | 81.7 | 84.1 | 86.5 | 89.0 | 91.4 |
| Italy | 84.7 | 85.9 | 87.0 | 88.1 | 89.2 | 90.3 | 91.4 | 92.6 | 93.7 | 94.8 |
| Japan | 59.3 | 63.5 | 67.6 | 71.8 | 75.9 | 80.0 | 84.2 | 88.3 | 92.5 | 96.6 |
| Latvia | 46.8 | 53.5 | 60.2 | 66.9 | 73.6 | 80.3 | 87.0 | 93.7 | 100.4 | 107.1 |
| Liechtenstein | 72.6 | 76.6 | 80.6 | 84.6 | 88.6 | 92.6 | 96.6 | 100.6 | 104.6 | 108.6 |
| Lithuania | 90.2 | 91.7 | 93.2 | 94.8 | 96.3 | 97.8 | 99.4 | 100.9 | 102.4 | 104.0 |
| Luxembourg | 74.8 | 74.1 | 73.3 | 72.6 | 71.8 | 71.1 | 70.4 | 69.6 | 68.9 | 68.1 |
| Malta | 52.0 | 57.6 | 63.1 | 68.7 | 74.3 | 79.9 | 85.5 | 91.0 | 96.6 | 102.2 |
| Netherlands | 89.2 | 93.0 | 96.8 | 100.6 | 104.4 | 108.3 | 112.1 | 115.9 | 119.7 | 123.5 |
| Norway | 89.1 | 91.7 | 94.4 | 97.1 | 99.8 | 102.5 | 105.1 | 107.8 | 110.5 | 113.2 |
| Poland | -24.9 | -33.2 | -41.6 | -49.9 | -58.3 | -66.6 | -75.0 | -83.3 | -91.6 | -100.0 |
| Portugal | 98.1 | 100.9 | 103.7 | 106.5 | 109.3 | 112.1 | 115.0 | 117.8 | 120.6 | 123.4 |
| Romania | -187.2 | -179.7 | -172.2 | -164.6 | -157.1 | -149.5 | -142.0 | -134.4 | -126.9 | -119.4 |
| Slovakia | 137.8 | 135.8 | 133.8 | 131.8 | 129.8 | 127.8 | 125.8 | 123.8 | 121.8 | 119.8 |
| Slovenia | 106.4 | 106.1 | 105.9 | 105.7 | 105.4 | 105.2 | 104.9 | 104.7 | 104.5 | 104.2 |
| Spain | 92.7 | 95.8 | 99.0 | 102.1 | 105.3 | 108.5 | 111.6 | 114.8 | 117.9 | 121.1 |
| Sweden | 92.7 | 92.0 | 91.3 | 90.6 | 89.9 | 89.2 | 88.5 | 87.8 | 87.0 | 86.3 |
| Switzerland | 72.6 | 76.6 | 80.6 | 84.6 | 88.6 | 92.6 | 96.6 | 100.6 | 104.6 | 108.6 |
| United Kingdom | 120.3 | 113.1 | 105.8 | 98.6 | 91.3 | 84.1 | 76.9 | 69.6 | 62.4 | 55.1 |
| United States | 149.1 | 134.7 | 120.3 | 106.0 | 91.6 | 77.2 | 62.8 | 48.4 | 34.0 | 19.6 |

Source: ESRB(2022) and EIOPA(2022).

## 3.3 Downscaling to individual exposures

To improve the granularity of risk analyses, we consider the climate vulnerability of individual counterparties and financial risk metrics specifically retrieved for each ISIN-level exposure. Regarding counterparties, Figure 2 summarizes the carbon intensity data of the 25 economic segments considered in our study[7].

**Figure 2: C02 intensity by economic segment**

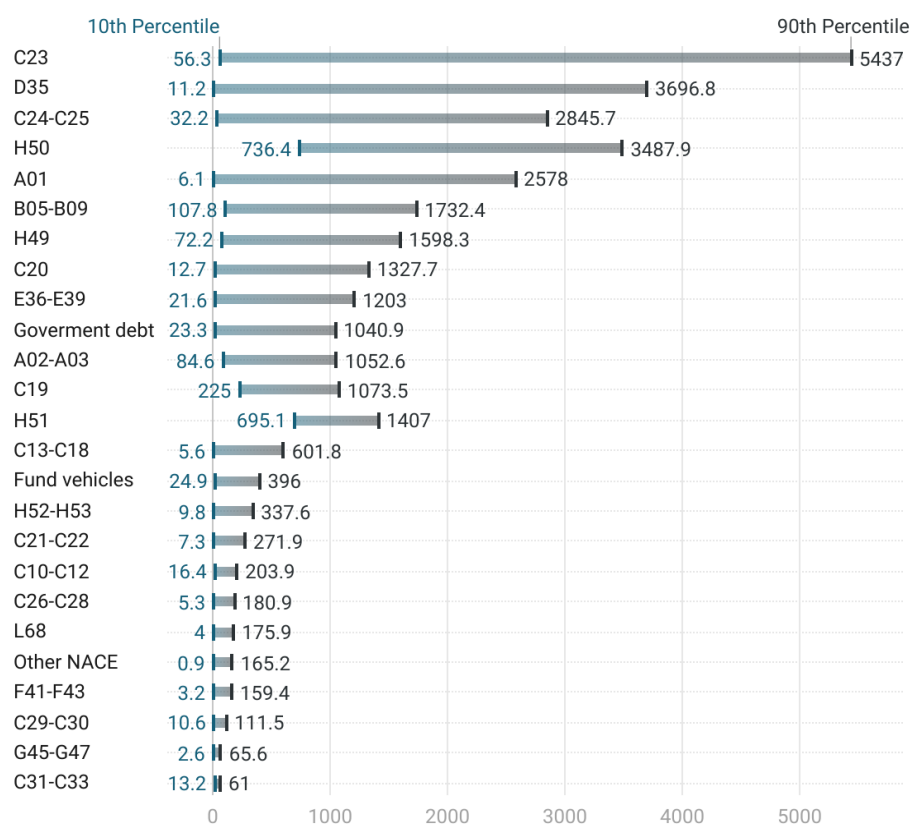

Note: Carbon intensity is calculated as the total direct (scope1) and indirect (scope 2) CO2-equivalent emissions in tones normalized by net sales or revenue in million US dollars.

Figure 2 shows that carbon intensity varies substantially across economic industries. The highest carbon intensity is observed in the C23 (manufacture of other non-metallic products), D35 (electricity, gas, stem and air conditioning supply) and H50 (water transportation) sectors. In contrast, information technology and professional, scientific and technical services (both included in other NACE) are among the lower emitters. Furthermore, the intra-sector dispersion of GHG emissions in many economic segments is remarkably high. For instance, the 10th to 90th quantiles of the C23 subsector range from 56.3 to 5,437 tC02eq/m$, whereas in the D35 subsector vary from 11.2 to 3,696.8.

---

[7] Carbon intensity data is defined as the total direct (scope1) and indirect (scope 2) CO2-equivalent emissions in tones normalized by net sales or revenue in million US dollars (tC02e/m$). For sovereign debt, carbon intensity data is retrieved as tC02e/GDP. For comparative purposes, the carbon intensity of sovereign countries is shown in the tC02e/m$ scale through a mapping of the quantile that each country occupies in the distribution for sovereign issuers to the corresponding quantile in the overall tC02e/m$ distribution.

To account for intra-sector GHG dispersion, we employ company-level data to differentiate among the best and worst-in-class in each economic industry. In the C23 subsector, the median carbon intensity is 891.8 tCO2e/m$, but there are firms with a carbon intensity lower than 100, whereas others show figures higher than 5,000. The GHG variability can be explained by the different technologies and energy sources that specific companies use to manufacture non-metallic products. The C23 NACE include all manufacturers of cement, glass, clay and ceramic products, regardless of whether their manufacturing process employs technologies based on renewable energies or traditional methods that burn fossil fuels. As a result, the industry median may significantly under or overestimate the carbon intensity of individual companies, leading to biases in the transition risk assessment. The relevance of company-level data is observed even when granular 4-digit NACE codes are employed, as explained in section 4.4.

To discriminate among individual firms, we employ a lognormal distribution to model the carbon intensity of economic industries. The lognormal distribution is a good candidate to describe carbon intensity data as: (i) GHG emissions can be assumed to be bounded by zero[8] and (ii) high-carbon emitters typically exhibit much higher carbon footprints than their sectoral peers, generating right-skewed distributions[9]. Furthermore, we find that the lognormal distribution provides an increasingly good fit for carbon intensity as the number of observations increase, as expected in a well-specified model. Figure 3 shows the Q-Q plot of C02 intensity for our set of 4,621 counterparties, showing that a lognormal model appropriately describes carbon intensity data ($R^2$ = 0.9851)[10].

**Figure 3: Lognormal Q-Q plot of CO2 intensity**

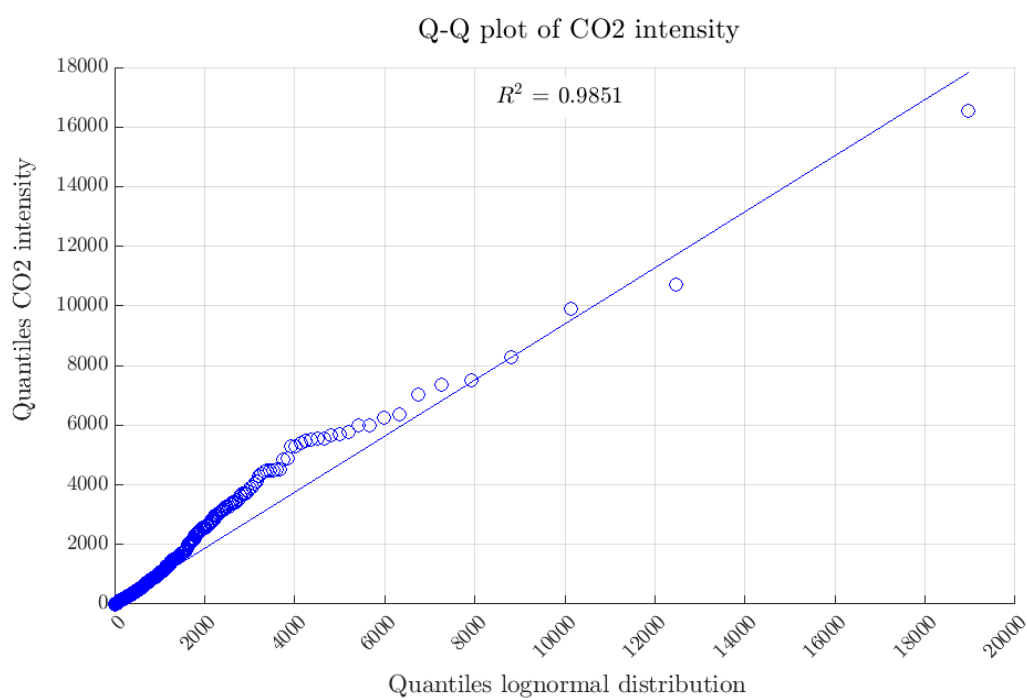

---

[8] Carbon offset can potentially generate negative carbon intensities. However, from a risk perspective, carbon offsets do not necessarily decrease the transition risk exposure of counterparties.
[9] Carbon intensity data exhibit positive skewness in 24 of our 25 economic segments.
[10] See Filliben (1975)

Regarding economic sectors, table 7 shows the carbon intensity mean and standard deviation of the 25 economic segments considered in our study. We calculate the raw carbon intensity mean $m_j$ and variance $v_j$ from the unique companies in each economic segment whereas the calibrated lognormal parameters $\mu$ and $\sigma$ are obtained as:

$$\mu = ln\left(\frac{m^2}{\sqrt{v+m^2}}\right) \text{ and } \sigma = \sqrt{log\left(\frac{v}{m^2}+1\right)}$$

The average $R^2$ of our sectoral Q-Q plots is 0.920, ranging from 0.665 to 0.995. The $R^2$ statistic is positively correlated with the number of observations, indicating that the goodness-of-fit improves with sample size. These sectoral distributions are employed to discriminate among the best and worst-in-class in each economic industry, as explained in Chapter 2.

Table 7: Carbon intensity distribution by economic segment

| NACE code / Asset class | No of firms | CO2 intensity | | CO2 distribution (lognormal) | | $R^2$ |
|---|---|---|---|---|---|---|
| | | mean | std | ln mean | ln std | |
| A01 | 10 | 806.6 | 1,066.1 | 6.19 | 1.01 | 0.7542 |
| A02-A03 | 3 | 431.0 | 539.5 | 5.59 | 0.97 | 0.6651 |
| B05-B09 | 141 | 804.2 | 1,063.4 | 6.18 | 1.01 | 0.9720 |
| C10-C12 | 107 | 98.5 | 122.4 | 4.12 | 0.97 | 0.9800 |
| C13-C18 | 70 | 237.3 | 344.4 | 4.90 | 1.06 | 0.9713 |
| C19 | 29 | 581.6 | 421.0 | 6.16 | 0.65 | 0.9741 |
| C20 | 96 | 478.2 | 743.7 | 5.56 | 1.11 | 0.9859 |
| C21-C22 | 91 | 88.7 | 135.9 | 3.88 | 1.10 | 0.9878 |
| C23 | 32 | 1,943.8 | 2,128.5 | 7.18 | 0.89 | 0.8684 |
| C24-C25 | 65 | 1,023.0 | 1,301.9 | 6.45 | 0.98 | 0.9376 |
| C26-C28 | 304 | 108.1 | 423.6 | 3.28 | 1.67 | 0.9420 |
| C29-C30 | 98 | 51.7 | 70.2 | 3.42 | 1.02 | 0.9665 |
| C31-C33 | 5 | 33.4 | 22.2 | 3.32 | 0.61 | 0.7622 |
| D35 | 126 | 1,431.7 | 2,405.5 | 6.60 | 1.16 | 0.9856 |
| E36-E39 | 21 | 519.6 | 480.3 | 5.94 | 0.79 | 0.7509 |
| F41-F43 | 84 | 88.1 | 264.6 | 3.33 | 1.52 | 0.9330 |
| G45-G47 | 164 | 53.4 | 135.9 | 2.97 | 1.42 | 0.9228 |
| H49 | 35 | 673.8 | 1,013.3 | 5.92 | 1.09 | 0.9718 |
| H50 | 24 | 2,000.8 | 1,445.6 | 7.39 | 0.65 | 0.9496 |
| H51 | 14 | 1,096.1 | 253.6 | 6.97 | 0.23 | 0.9237 |
| H52-H53 | 42 | 177.8 | 410.8 | 4.26 | 1.36 | 0.9278 |
| L68 | 124 | 85.9 | 135.3 | 3.83 | 1.12 | 0.9673 |
| Other | 819 | 101.4 | 454.5 | 3.10 | 1.70 | 0.9774 |
| Sovereign debt | 30 | 351.0 | 456.9 | 5.40 | 1.00 | 0.9253 |
| Other fund vehicles | 2087 | 193.8 | 268.3 | 4.70 | 1.03 | 0.9946 |
| All issuers | 4621 | 257.6 | 691.1 | 4.50 | 1.45 | 0.9851 |

Notes. The number of firms indicates the unique counterparties in each economic segment with available carbon intensity data. The $R^2$ statistic measures the goodness-of-fit (see Filliben, 1975). Carbon intensity is measured as the total scope 1 and scope 2 CO2-equivalent emissions in tones normalized by net sales or revenue in million US dollars.

Similarly, table 8 shows the summary statistics for the credit and market risk metrics obtained in our 25 economic segments. Following carbon intensity, credit and market risk metrics exhibit substantial intra-sector variability, motivating the use of ISIN-level data to complement sectoral analyses. These credit and market risk metrics are employed to estimate the potential loss that financial instruments with different credit ratings, duration, convexity or price volatility can experience upon occurrence of a transition risk scenario[11].

**Table 8: Market and credit risk metrics by economic segment**

| NACE Code / Asset Class | No of ISINs | Equity volatility | | | | CQS | | | | Spread duration | | | |
|---|---|---|---|---|---|---|---|---|---|---|---|---|---|
| | | p10 | median | mean | p90 | p10 | median | mean | p90 | p10 | median | mean | p90 |
| A01 | 26 | 16.0 | 29.9 | 31.9 | 51.3 | 2.00 | 2.50 | 2.50 | 3.00 | 0.06 | 1.80 | 2.51 | 6.44 |
| A02-A03 | 6 | 32.3 | 36.7 | 36.7 | 41.2 | 2.00 | 2.50 | 2.50 | 3.00 | 0.18 | 2.20 | 2.43 | 5.12 |
| B05-B09 | 258 | 27.9 | 43.7 | 48.1 | 67.5 | 2.00 | 2.00 | 2.53 | 4.00 | 1.36 | 3.50 | 4.48 | 8.57 |
| C10-C12 | 277 | 16.0 | 24.1 | 25.7 | 39.1 | 2.00 | 2.00 | 2.40 | 3.00 | 0.93 | 3.80 | 5.17 | 10.41 |
| C13-C18 | 137 | 21.7 | 30.5 | 33.1 | 47.0 | 2.00 | 2.00 | 2.51 | 3.00 | 0.60 | 3.70 | 4.04 | 7.56 |
| C19 | 93 | 22.7 | 30.3 | 32.4 | 46.3 | 2.00 | 2.00 | 2.33 | 3.00 | 0.82 | 4.40 | 6.75 | 19.26 |
| C20 | 192 | 19.2 | 27.8 | 41.9 | 45.0 | 2.00 | 2.00 | 2.35 | 3.00 | 0.57 | 3.90 | 5.78 | 15.72 |
| C21-C22 | 298 | 19.0 | 32.1 | 40.6 | 63.5 | 2.00 | 2.00 | 2.39 | 3.00 | 0.98 | 4.00 | 5.66 | 9.99 |
| C23 | 74 | 20.9 | 29.6 | 33.7 | 55.5 | 2.00 | 2.00 | 2.59 | 4.00 | 0.26 | 3.00 | 3.08 | 6.38 |
| C24-C25 | 128 | 24.2 | 33.7 | 37.3 | 53.3 | 2.00 | 2.00 | 2.39 | 3.00 | 0.22 | 2.10 | 2.39 | 5.32 |
| C26-C28 | 642 | 22.6 | 33.8 | 41.3 | 60.5 | 2.00 | 2.00 | 2.36 | 3.00 | 0.73 | 3.70 | 4.12 | 7.65 |
| C29-C30 | 288 | 22.9 | 32.1 | 36.6 | 56.2 | 2.00 | 2.00 | 2.47 | 3.00 | 0.37 | 3.40 | 3.71 | 7.51 |
| C31-C33 | 10 | 23.7 | 30.7 | 32.3 | 45.7 | 3.00 | 3.00 | 3.00 | 3.00 | 2.13 | 2.10 | 2.13 | 2.13 |
| D35 | 458 | 17.1 | 26.5 | 29.8 | 46.5 | 2.00 | 2.00 | 2.31 | 3.00 | 0.62 | 4.20 | 7.04 | 17.26 |
| E36-E39 | 56 | 16.1 | 27.6 | 27.4 | 41.5 | 2.00 | 2.00 | 2.42 | 3.00 | 0.08 | 1.30 | 3.35 | 7.20 |
| F41-F43 | 232 | 21.1 | 27.7 | 30.5 | 42.2 | 2.00 | 2.00 | 2.45 | 3.00 | 0.21 | 1.50 | 2.40 | 6.58 |
| G45-G47 | 417 | 20.8 | 32.1 | 36.5 | 56.4 | 2.00 | 2.00 | 2.42 | 3.00 | 0.57 | 3.60 | 3.85 | 7.57 |
| H49 | 102 | 19.4 | 27.5 | 29.2 | 40.8 | 2.00 | 2.00 | 2.43 | 3.00 | 0.50 | 3.40 | 4.29 | 9.93 |
| H50 | 56 | 32.2 | 48.3 | 48.3 | 61.7 | 2.00 | 2.00 | 2.76 | 4.80 | 0.73 | 2.50 | 3.48 | 7.72 |
| H51 | 48 | 32.2 | 39.1 | 43.0 | 47.9 | 2.00 | 2.00 | 2.98 | 6.00 | 1.42 | 3.70 | 4.29 | 6.70 |
| H52-H53 | 103 | 20.8 | 26.8 | 31.8 | 39.9 | 2.00 | 2.00 | 2.61 | 4.00 | 1.14 | 4.10 | 4.42 | 7.66 |
| L68 | 467 | 17.6 | 24.2 | 28.0 | 40.6 | 2.00 | 2.00 | 2.39 | 3.00 | 0.83 | 4.30 | 5.62 | 12.89 |
| Other NACE | 5,387 | 19.7 | 30.6 | 38.3 | 57.0 | 2.00 | 2.00 | 2.33 | 3.00 | 0.66 | 3.60 | 5.14 | 11.80 |
| Sovereign debt | 1,413 | - | - | - | - | 1.00 | 2.00 | 1.92 | 3.00 | 0.46 | 4.30 | 5.56 | 11.93 |

Notes. The number of ISINs indicates the unique financial instruments in each economic segment with at least one data field available (i.e.: volatility, CQS or duration).

---

[11] Given our large heterogeneous dataset (w. 7,060 unique bonds), we calculate duration and convexity through approximative formulas than can be applied to both fixed and floating rate exposures. Specifically, we estimate duration as $T / (1+c)^{(T/2)}$, where T is the time to maturity in years and c is the bond coupon, whereas convexity is approximated as $[T*(T+1)] / (1+c)^2$. For convexity, an upper bound of forty times the duration is employed to avoid unreasonable values in very long dated exposures.

# 4. Empirical results

This section presents the results from our transition risk analyses. We first examine the transition risk distribution for different asset classes. Next, we consider the distribution of MtM losses across investment fund portfolios, analyzing which vehicles suffer the highest and lowest losses. We also consider how sustainable funds perform in the low-carbon transition. Finally, we present the results from an alternative risk measure that is designed to tackle the comparability problems that are usually observed in the measurement of GHG emissions.

## 4.1 Transition risk losses by asset class

Figure 4 shows the distribution of transition risk losses by asset class. As expected, the highest losses are observed in equity investments (-12.71% on average), followed by corporate bonds (-5.61%) and sovereign debt (-4.77%). However, there is substantial variability across the financial instruments included in each asset class. Table 9 provides a characterization of the worst and best performers by asset class. The worst 1% equities suffer an average loss of 70.96% and are characterized by companies with a high carbon footprint (1,812.3 tC02e/m$ on average) operating in highly pollutant sectors (e.g.: NACE codes B, C19 and D35). On the other hand, equities issued by companies with near-zero carbon emissions are among the most resilient in our transition risk analyses.

In corporate debt, it stands out that a large proportion of bonds suffer small MtM losses. In addition to bonds issued by low-carbon companies, almost half of the corporate debt in our database exhibits a short maturity (less than 3 years), enduring small losses in a credit spread widening scenario. In contrast, corporate bonds issued by high emitter companies (755.24 tC02e/m$) with long residual maturities (24.01 average duration) suffer the largest losses.

**Figure 4: Distribution of transition risk losses by asset class**

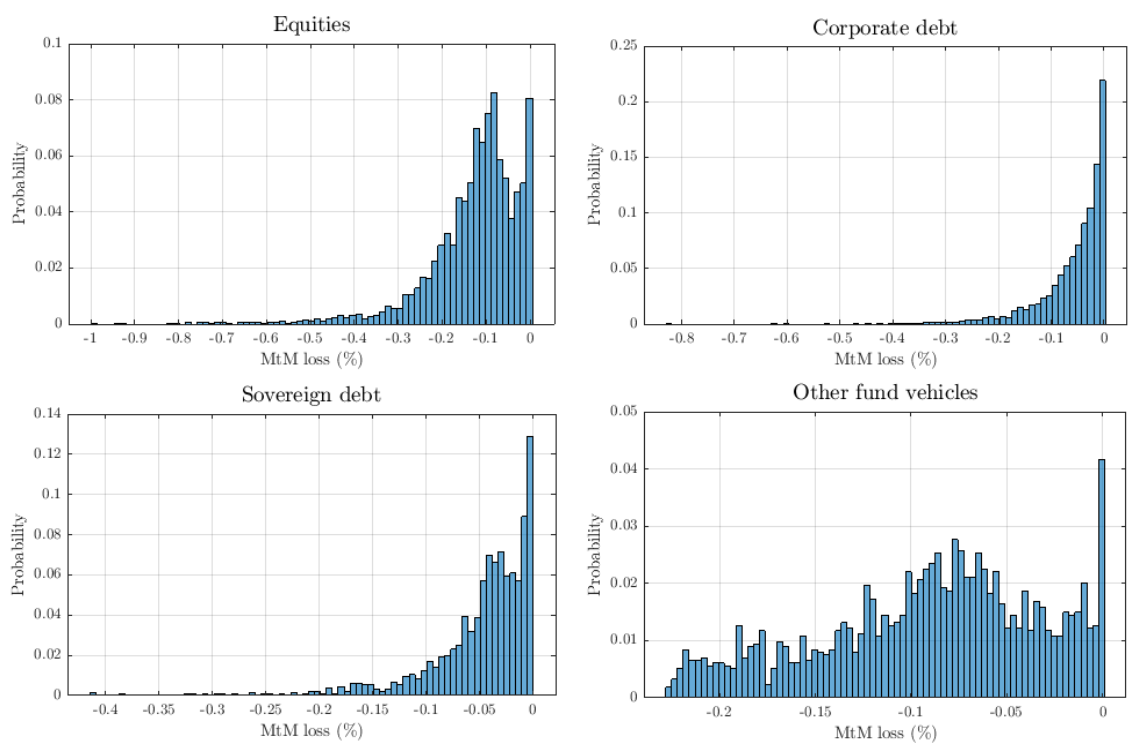

The transition risk distribution for sovereign debt also exhibits a high proportion of bonds with low MtM losses. Similar to corporate debt, the resilience of these instruments stems from the combined effect of low-carbon issuers and bonds with short-term maturities. Overall, two thirds of the sovereign bonds in our database suffer a MtM loss lower than 5%, outperforming other asset classes. However, sizable MtM losses are observed in very long dated sovereign bonds (32.5 duration) issued by countries that are particularly exposed to the climate transition.

To understand the results for sovereign debt, note that the macroeconomic scenario calibrated by the ECB and ESRB provides reference yield curve increase for the main sovereign issuers (see table 6). The largest yield increases are observed in long-dated bonds of several European countries. Therefore, long-term bonds issued by some EU countries are ranked among the worst performers despite having good credit ratings and relatively low carbon intensity. In contrast, the calibrated scenario assigns negative yield shocks to the debt issued by certain countries (e.g.: Poland, Romania and several maturity buckets in China and Hungary). As a result, credit exposures with negative yield shocks perform well despite being issued, particularly in the case of China, by a country that is the world's largest emitter of $CO_2$.

Finally, averaging portfolio effects and the use of representative holdings make the loss distribution for other fund vehicles more concentrated than in other asset classes. However, the distribution still provides notable discrimination among different portfolios. The worst performing vehicles are pure equity funds that invest in companies with a high carbon footprint (1,127.28 tC02e/m$ on average). In contrast, funds investing in corporate and government bonds issued by low GHG emitters are among the best performers.

**Table 9: Characterization of worst and best financial instruments by asset class**

| Asset class | MtM loss (%) | Climate vulnerability | | Credit and market risk | | | |
|---|---|---|---|---|---|---|---|
| | | Carbon intensity (tC02e/m$) | NACE code / country | CQS | Duration (years) | Volatility (%) | Investment style |
| Equities (worst 1%) | -70.96 | 1,812.3 | B05-09; C19; D35 | - | - | 66.33 | - |
| Corporate bonds (worst 1%) | -37.22 | 755.24 | D35, others | 2.72 | 24.01 | - | - |
| Sovereign debt (worst 1%) | -31.03 | 59.54 | BE, DE, FR, ES | 1.23 | 32.54 | - | - |
| Other funds (worst 1%) | -22.20 | 1,127.28 | - | - | - | - | Equities |
| Equities (best 1%) | <-0.01 | 5.89 | A01; C21,22,24-28; L68 | - | - | 34.94 | - |
| Corporate bonds (best 1%) | <-0.01 | 7.88 | C13-18; C23-25 | 2.09 | 3.72 | - | - |
| Sovereign debt (best 1%) | <-0.01 | 116.83 | PO, RO, HU, CH | 1.82 | 0.12 | - | - |
| Other funds (best 1%) | <-0.01 | 2.46 | - | - | - | - | Corporate & Gov. bonds |
| Equities (all) | -12.71 | 271.79 | - | - | - | 37.06 | - |
| Corporate bonds (all) | -5.61 | 190.73 | - | 2.27 | 4.52 | - | - |
| Sovereign debt (all) | -4.77 | 351.0 | - | 1.92 | 5.31 | - | - |
| Other funds (all) | -9.07 | 193.8 | - | - | - | - | - |

Notes: MtM loss, carbon intensity, CQS, duration and volatility figures are expressed as the equal-weighted average across all the financial instruments included in the corresponding subset.

**4.2 Transition risk losses in Spanish investment funds**

Figure 5 shows the distribution of transition risk losses in Spanish funds. The average loss in the investment fund sector is 5.69%, which amounts to a total MtM loss of EUR 17.5 billion[12]. This loss only considers the direct, first-round effects of the climate transition. Amplifying factors like the fund-flow relation, market impact, indirect contagion or other system-wide drivers could trigger feedback loops and non-linear effects that significantly increase the final loss.[13]

The transition risk distribution for investment funds is significantly left-skewed (-0.97 skewness), showing notable dispersion across investment portfolios. In a disorderly transition scenario, the worse 1% funds suffer a loss of 21.33% on average, whereas the best 1% funds do not experience any MtM loss. The detailed ISIN-level composition of each investment fund allows us to analyze the drivers of transition risk and characterize the portfolios that suffer the highest and lowest MtM losses.

**Figure 5: Transition risk distribution of the investment fund sector**

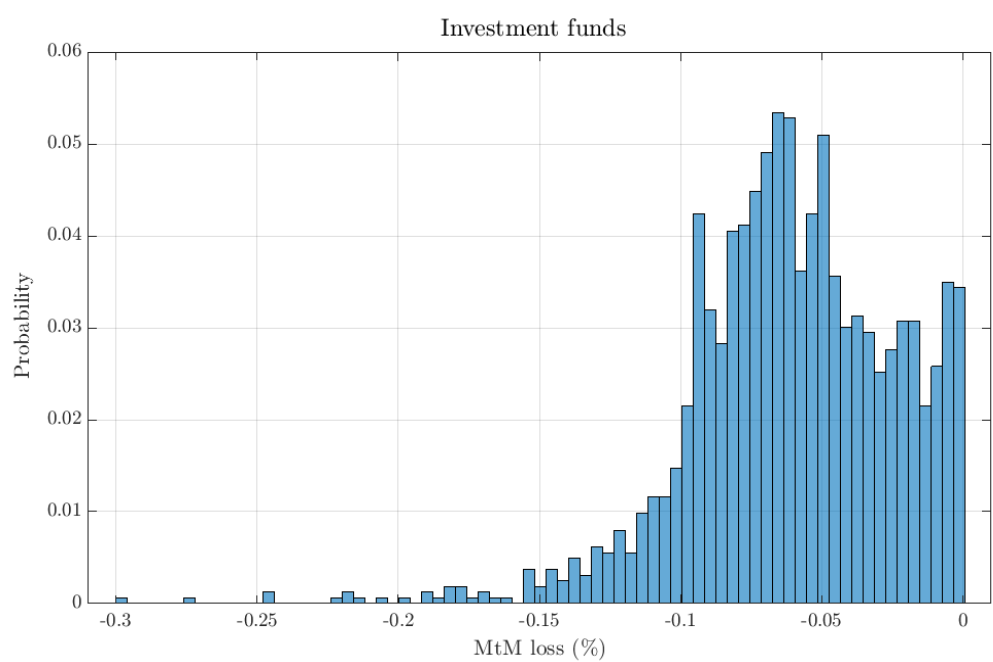

Funds investing in equities of highly pollutant companies exhibit the worst performance in the climate transition. Specifically, the worst 1% funds invest 94.8% of its portfolio in equities and exhibit an AuM-weighted carbon intensity of 998.9 tCO2e/m$[14]. In contrast, the aggregate fund sector only invests 15.46% in equities and shows a carbon intensity of 137.2. Figure 6 presents the sectoral breakdown of the 10 investment funds with the highest MtM loss. The 10 worst performing funds invest 64.6% of their portfolio in the Climate Policy Relevant Sectors (CPRS) that are expected to perform worse in the

---

[12] The MtM loss increase to EUR 820.5 billion if extrapolated to the European fund sector.
[13] Management fees, depositary fees and other expenses are not included in the calculation as they are not related to the climate transition. These costs will increase the loss experienced by fund investors. In addition, the dynamic behavior of fund managers in response to transition risk could also increase or reduce the MtM loss.
[14] Carbon intensity figures for investment fund portfolios are calculated as the AuM-weighted carbon intensity of all individual fund positions. These AuM-weighted figures are hence not directly comparable to the equal-weighted averages included in tables 7 and 9.

low-carbon transition (see Battiston et al., 2017)[15]. In comparison, the overall fund sector only held 12.2% of its portfolio in CPRS sectors.

Conversely, there are 34 out of 1,626 funds that do not experience any MtM loss. However, this figure is partially driven by funds that are either in liquidation or recently formed. In particular, half of the funds with no MtM loss have an AuM lower than the legal minimum, indicating that these vehicles are in constitution or liquidation. Disregarding these vehicles, the best 1% funds maintain most of their portfolio in cash and cash-equivalent instruments, enduring a low MtM loss in an adverse climate scenario.

**Figure 6: Sectoral breakdown of the 10 worst performing funds**

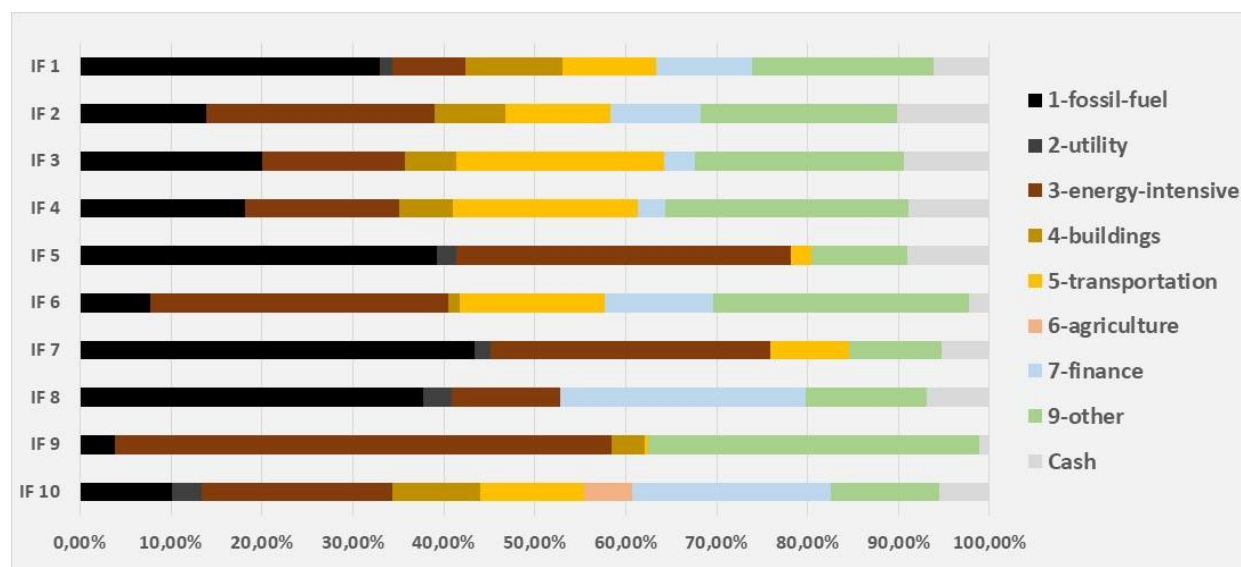

### 4.3 Sustainable funds

Sustainable fund portfolios exhibit significant differences with the aggregate fund sector in terms of transition risk[16]. As table 10 shows, sustainable funds held a higher share of equities than traditional funds (25.17% vs 15.46% in the fund sector). However, even overweighing the riskier asset class, sustainable funds outperform the overall fund sector in the climate transition. Regarding tail risk, the worst 1% and 5% sustainable funds experience an average loss of 14.65% and 11.00% (vs. 21.34% and 15.47% in the fund sector). Similarly, the average MtM loss across all sustainable funds is 5.70%, which outperform the 5.92% loss that would be obtained in the fund sector portfolio with a comparable allocation in terms of asset classes.

The portfolio of sustainable funds also exhibits lower carbon intensity than the overall fund sector (115.52 vs 137.22). The smaller C02 footprint is driven by both: (i) an overweighing of low-carbon sectors and (ii) a selection of financial counterparties that exhibit lower $CO_2$ intensity than their sectoral peers. However, table 10 shows that the climate outperformance of sustainable funds is not homogeneous across asset classes. In equities and other fund vehicles, sustainable fund invests in instruments with lower C02 footprint that outperform their sectoral peers in the low-carbon transition.

---

[15] The CPRS sectors are fossil fuels, utilities, energy intensity, buildings, and agriculture.
[16] The subset of sustainable funds in the Spanish fund sector is obtained from Cambón and Ispierto (2022), which define sustainable funds as those complying with 2014 the Inverco SRI Circular.

For instance, the equity portfolio of sustainable funds shows an AuM-weighed carbon intensity of 147.40 and endure a MtM loss of 7.82%. In comparison, the equity portfolio of the entire fund sector exhibits a carbon intensity of 224.37 and suffer a MtM loss of 9.30%[17].

In contrast, the corporate bond portfolio of sustainable funds exhibits a slightly higher carbon intensity than in the aggregate sector. Consequently, the MtM loss experienced by sustainable funds in corporate bonds is higher than the loss observed in the overall fund sector (-5,57% vs. -4.02%). This finding suggests that the investment choices of sustainable funds in corporate bonds could be further improved to obtain a more climate-friendly portfolio.

**Table 10: Sustainable funds performance by asset class**

| Asset class | Sustainable funds | | | Overall fund sector | | |
|---|---|---|---|---|---|---|
| | Investment share (% AuM) | MtM loss (%) | Carbon intensity (tC02e/m$) | Investment share (% AuM) | MtM loss (%) | Carbon intensity (tC02e/m$) |
| Equities | 25.17% | -7.82 | 147.40 | 15.46% | -9.30 | 224.37 |
| Corporate bonds | 24.17% | -5.57 | 143.30 | 19.68% | -4.02 | 137.43 |
| Sovereign debt | 15.49% | -4.68 | 56.46 | 20.97% | -3.27 | 65.18 |
| Other funds | 26.03% | -6.38 | 134.63 | 34.42% | -8.06 | 179.59 |
| Entire portfolio | 100.00% | -5.70 | 115.52 | 100.00% | -5.69 / -5.92 | 137.22 |
| Worst 1% funds | - | -14.65 | 201.03 | - | -21.34 | 998.92 |

Notes: MtM losses are calculated as the AuM-weighted average across all positions in the corresponding portfolio. The -5.92% loss in the overall fund sector represents the MtM loss that would be obtained in the fund sector portfolio with an asset class allocation equivalent to that of sustainable funds.

**4.4 Robustness check: Alessi and Battiston (2022)**

A notable challenge for climate risk assessments is the lack of comparable and independently verified data on GHG emissions. Given the problems observed in GHG measurement[18], Alessi and Battiston (2022) develops two sectoral metrics of greenness and transition risk that are transparent and easily replicable. Using 4-digit NACE codes, Alessi and Battiston (2022) quantifies greenness as the share of each economic sector that is aligned with the EU taxonomy for sustainable activities (Taxonomy Alignment Coefficient or TAC). In addition, since greenness does not provide a direct risk assessment, Alessi and Battiston (2022) also considers the share of each sector that is invested in high-carbon activities (Transition-risk Exposure Coefficient or TEC).

Table 11 exhibits the TAC and TEC of Spanish funds compared with the EU sector. To understand these figures, note that Alessi and Battiston (2022) applies only to investment position with a NACE code (i.e.: equity and corporate bonds). Consequently, exposures without a NACE code (i.e.: sovereign debt and other fund vehicles) are counted as zero in the portfolio aggregation, driving down the overall TAC and TEC. To complement these metrics, we also compute an adjusted TAC and TEC by considering the share of each portfolio that is included in the calculation. This adjustment provides standardized figures that can be used to compare investment portfolios with notably different compositions.

---

[17] The MtM loss figures reported for investment funds are calculated as the AuM-weighted loss across all positions in the corresponding portfolio. These AuM-weighted figures are hence not directly comparable to the equal-weighted averages included in table 9. For instance, the equal-weighted loss across all equity instruments is -12.70%, which is higher than the AuM-weighted loss obtained by the equity portfolio of sustainable funds (-7.82%) and the AuM-weighted loss obtained by the equity portfolio of the fund universe (-9.30%).

[18] While GHG reporting has improved in recent years, data quality issues hinder the comparability of different climate studies, leading to diverging outcomes depending on the data sources (see for instance Bingler et al. (2022) and chapter 32 of NGFS, 2020).

Table 11: Taxonomy alignment and transition risk exposure of investment funds

| Holder | Transition-risk Exposure (TEC) | Taxonomy Alignment (TAC) | Eligible portfolio | Adjusted TEC | Adjusted TAC |
|---|---|---|---|---|---|
| ES investment funds | 4.37% | 0.94% | 33.88% | 12.91% | 2.79% |
| ES sustainable funds | 3.78% | 2.67% | 47.98% | 7.87% | 5.57% |
| EU investment funds | 6.11% | 1.37% | 20.91% | 29.20% | 6.54% |

Note: The TAC and TEC of EU investment funds are retrieved from Alessi and Battiston (2021). Adjusted TAC and TEC values are calculated as the standard TAC and TEC divided by the eligible portfolio.

Table 11 show that Spanish funds are less exposed to transition risk than their EU counterparts, but also exhibit a lower alignment with the EU taxonomy. The difference can be explained by the relation that TAC and TEC show in many economic sectors. Specifically, of all 4-digit NACE codes exhibiting a positive TEC, roughly half of them also exhibit a positive TAC. Therefore, funds investing in sectors with high transition risk also tend to exhibit a higher alignment with the EU taxonomy[19].

As a representative example, the TEC of NACE sector 35.11 (production of electricity) is 0.39, which is the share of electric energy that is derived from fossil fuels. However, this sector also presents a TAC of 0.35, which correspond to the share of electricity that is derived for renewable sources. Consequently, all companies in NACE 35.11 receive high TEC and TAC values, irrespectively of whether their electricity is generated through renewable sources or by burning fossil fuels. This example, although arguably not applicable to all sectors, illustrates how even 4-digit sectoral breakdowns can be problematic when assessing transition risk, motivating the use of company-level data to supplement sectoral analyses.

Finally, table 11 shows that the portfolio of sustainable funds is greener and less exposed to transition risk than the Spanish and EU fund sectors. Given the sectoral focus of Alessi and Battiston (2022), this finding suggests that sustainable funds avoid economic sectors that are highly exposed to transition risk and invest a higher share of their portfolio in industries that are aligned with the EU taxonomy.

---

[19] The correlation between the TEC and TAC figures in our sample of 1629 funds is 0.55

# 5. Conclusion

This paper proposes a comprehensive methodology to analyze the vulnerability of investment portfolios to the low-carbon transition. To measure transition risk, we combine climate-related metrics obtained for each counterparty with traditional measures of credit and market risk. This methodology allows us to quantify the loss of value that each individual exposure, and hence the corresponding portfolio, could suffer upon materialization of a transition risk scenario.

We find that investment funds suffer a moderate 5.7% loss in a high transition risk scenario. However, the transition risk distribution is significantly left-skewed, with the 1% worst performing funds experiencing an average loss of 21.3%. In terms of asset classes, equities are the worst performers (-12.7%), followed by corporate bonds (-5.6%) and sovereign debt (-4.8%). We also find that sustainable funds are less exposed to transition risk and perform better that the overall fund sector in the low-carbon transition, validating their choice as green investments. In addition, we show that Spanish funds also exhibit lower transition risk that their EU counterparts.

Regarding future work, the inclusion of scope 3 emissions in the transition risk analyses could further improve the discrimination across different portfolios, economic sectors and individual counterparties. However, availability and quality issues in the reporting of scope 3 emissions currently hinder their systematic use. Similarly, while Spanish funds make limited use of derivatives, quantifying the risk exposure obtained through financial derivatives (which may affect both market and counterparty risk) could also improve the climate risk assessment[20].

Finally, while the focus of this paper is on transition risks, the relation between the costs of the transition and future physical risks should be considered in the risk assessment. Specifically, although the climate transition will generate substantial costs for companies, these risks should be appropriately weighed against the benefits of limiting global warming to evaluate the costs and opportunities arising from a low-carbon economy.

---

[20] In addition, our monitoring framework could be also enriched by considering a look-through approach for other fund vehicles.